\documentclass[prl,twocolumn,showpacs,superscriptaddress]{revtex4}
\usepackage{graphicx,color,amsmath,amssymb}
\usepackage[colorlinks=true, linkcolor=blue]{hyperref}

\def\reals{\mathbb{R}}

\def\({\left(}
\def\){\right)}

\renewcommand{\vec}[1]{\mathbf{#1}}
\newcommand{\mat}[1]{\mathsf{#1}}

\begin{document}

\title{One-Way Quantum Computing in the Optical Frequency Comb}
\author{Nicolas C. Menicucci}
\email{nmen@princeton.edu}
\affiliation{Department of Physics, Princeton University, Princeton, New Jersey 08544, USA}
\affiliation{Department of Physics, University of Queensland, Brisbane, Queensland 4072, Australia}
\author{Steven T. Flammia}
\email{sflammia@perimeterinstitute.ca}
\affiliation{Perimeter Institute for Theoretical Physics, Waterloo, Ontario, N2L 2Y5 Canada} 
\author{Olivier Pfister}
\email{opfister@virginia.edu}
\affiliation{Department of Physics, University of Virginia, Charlottesville, Virginia 22903, USA}

\pacs{03.67.Lx, 03.67.Mn, 42.50.Dv, 42.65.Yj}

\date{\today}


\begin{abstract}
One-way quantum computing allows any quantum algorithm to be implemented easily using just measurements.  The difficult part is creating the universal resource, a cluster state, on which the measurements are made.  We propose a radically new approach: a scalable method that uses a single, multimode optical parametric oscillator (OPO).  The method is very efficient and generates a continuous-variable cluster state, universal for quantum computation, with quantum information encoded in the quadratures of the optical frequency comb of the OPO.
\end{abstract}

\maketitle

\paragraph{Introduction}

Quantum computing (QC) is a fascinating endeavor, ripe with promises of exponential speedup of particular mathematical problems such as quantum system simulation~\cite{Feynman1982} and integer factoring~\cite{Shor1994}. Because implementing QC requires exquisite control of each single quantum memory unit (e.g., qubit) in a large-size register, practical QC is therefore faced with the daunting challenges of overcoming decoherence and achieving scalability~\cite{Braunstein2001}. Recently, the invention of one-way quantum computing introduced a new paradigm for quantum information processing~\cite{Raussendorf2001} based on teleportation alone~\cite{Gottesman1999}.  In the traditional ``circuit" QC model~\cite{Nielsen2000}, physical quantum systems carry quantum information and undergo controlled unitary evolution; in the one-way QC model, quantum information exists virtually in a {\em cluster state}~\cite{Briegel2001} and is manipulated through a sequence of local measurements. The choice of measurement basis and the measurement results fully determine the quantum algorithm.

The appeal of one-way QC is that it consolidates most of the challenging work into creating the universal resource---the cluster state---and that it only requires local measurements.  In addition, some one-way QC schemes admit very high fault-tolerance thresholds~\cite{Dawson2006}, and experimental realizations with four qubits have already been achieved~\cite{Walther2005,Tame2007}.  Efficient methods of creating large-scale cluster states are still needed, however, for practical implementation to be realistic.  Here, we describe a radically new approach to scalability: a ``top-down" method to produce large continuous-variable~(CV) cluster states using a compact experimental setup. The interest of CVs is their natural implementation by squeezing (quantum noise reduction) in quantum optical systems. Photons are also less prone to decoherence than, say, atoms, due to their lower propensity to interact with the environment. Several studies have established the use of photonic CVs for teleportation~\cite{Furusawa1998,Yonezawa2004}, QC~\cite{Lloyd1999}, quantum error correction~\cite{Braunstein1998a}, cluster states~\cite{Zhang2006,Menicucci2006,Menicucci2007,Su2007}, and one-way QC~\cite{Menicucci2006}.

\paragraph{CV cluster states from a single OPO}

Our implementation of a scalable one-way quantum processor uses one multimode optical parametric oscillator (OPO). Initial proposals for constructing CV cluster states~\cite{Zhang2006,Menicucci2006} involved inline squeezers (seeded OPOs)~\cite{Yurke1985}, which are difficult to implement.  A more viable method relies on the Bloch-Messiah decomposition~\cite{Braunstein2005} and uses $N$~vacuum squeezers followed by an $\mathcal{O}(N^2)$-port interferometer~\cite{vanLoock2007}.  Our method improves further, requiring only a {\em single OPO\/} and {\em no interferometer\/}~\cite{Menicucci2007}, and provides huge scaling potential.

The OPO combines two essential elements. The first one is an optical cavity (e.g., two facing mirrors) whose spectrum of resonant frequencies forms an optical frequency comb~(OFC), so called because of the equal spacing between modes. Considered as a quantum system, the OFC is a large collection of independent quantum harmonic oscillator modes, or ``qumodes'' (a term used in analogy with ``qubit''). Quantum information is encoded  in the quadrature field variables of the OFC, which are analogs of position and momentum for a mechanical oscillator. The OFC can comprise millions of modes and has outstanding classical coherence properties that have found groundbreaking applications in the revolutionary and now ubiquitous use of mode- and carrier-envelope-phase-locked femtosecond lasers in time/frequency metrology~\cite{Hall2006,Hansch2006}.

The second crucial element of the OPO is the nondissipative nonlinear medium placed in the cavity. In a basic OPO, this medium is pumped by a monochromatic field and promotes downconversion, i.e.,~the simultaneous absorption of a pump photon at frequency~$\omega_\text{p}$ and emission of a photon pair at OFC frequencies~$\omega_m$ and~$\omega_n$ such that $\omega_ \text{p}=\omega_m+\omega_n$, as well as upconversion, the reverse process.  Such interactions yield bipartite CV entanglement of the OFC qumodes at frequencies~$\omega_m$ and~$\omega_n$~\cite{Braunstein2005a}. In the OPO we propose to use, the nonlinear medium is specifically engineered to quasiphasematch~\cite{Armstrong1962,Fejer1992} several such interactions simultaneously~\cite{Pooser2005}. This, along with a polychromatic pump, allows one to ``write" an {\em entangled network} onto the OFC using pairwise couplings.  We show in this paper that this network can be made to precisely constitute a large CV cluster state, universal for one-way QC, and whose scaling 
requires no increase in pump complexity (number of frequencies) and only {\em linear} increases in pump intensity and nonlinear gain bandwidth.  

A CV cluster state~\cite{Zhang2006,Menicucci2006} is defined as any member of a family of squeezed states indexed by an overall squeezing parameter $r>0$ for which the variance of each component of $(\vec p - \mat A \vec q)$ tends to~0 as $r \to \infty$~\cite{vanLoock2007}.  Here, $\vec q = (q_1, \dotsc, q_N)^T$ and $\vec p = (p_1, \dotsc, p_N)^T$ are vectors of amplitude and phase quadrature operators, respectively, and $\mat A$ is the (weighted, undirected) adjacency matrix of the cluster state's $N$-node graph~\cite{Menicucci2007}. The infinite-squeezing limit is not achievable by any finite-energy state, an important point we address later.  

We consider an OPO that implements the following Hamiltonian, in the interaction picture and assuming a classical undepleted pump~\cite{Pfister2004,Menicucci2007}:
\begin{align}
\label{eq:Hamiltonian}
	\mathcal{H}(\mat A) = i \hbar \kappa \sum_{m,n} A_{mn} (a^\dag_m a^\dag_n - a_m a_n)\;,
\end{align}
where $\kappa>0$ is an overall nonlinear coupling strength (squeezing parameter per unit time). The nodes of the graph described by $\mat A$ correspond to OPO qumodes and any ($m$,$n$) edge is weighted by $A_{mn}$, whose magnitude is the qumode coupling strength in units of~$\kappa$ and whose sign indicates downconversion if positive and upconversion if negative. In previous work~\cite{Menicucci2007}, we showed that $\mathcal{H} (\mat A)$ always generates a CV cluster state whose graph is generally not given by  $\mat A$.  If $\mat A$ is an \emph{orthogonal} matrix, though, the graph \emph{is} given by $\mat A$---a fact we previously used to construct an $\mathcal H (\mat A)$ that generates large sets of very small ($2\times 2$ or $2\times 3$) CV cluster states~\cite{Zaidi2008}.  Being disconnected, these cluster states are not universal for QC, but requiring orthogonality is a useful simplification.  Here we construct an orthogonal $\mat A$ for which $\mathcal H (\mat A)$ generates a QC-universal CV cluster state.

Labeling rows and columns of~$\mat A$ by sequential OFC modes, $\mathcal H (\mat A)$ is easy to implement experimentally when $\mat A$ is a \emph{Hankel} matrix (i.e., has constant skew diagonals).  Any pump frequency~$\omega_\text{p}$ satisfying $\omega_\text{p} = \omega_m + \omega_n$, together with the assumption of a constant interaction strength, generates a constant skew diagonal in $\mat A$ and sets all its entries to the same value, fixed by the pump power.  Additional pump frequencies generate additional skew diagonals, resulting in $\mat A$ having Hankel form.  This connection, along with a useful shorthand for Hankel matrices, is illustrated in Fig.~\ref{fig:expt} and in the supplemental movie \cite{movie}.

\begin{figure}
\begin{center}
\includegraphics[width=.75\columnwidth]{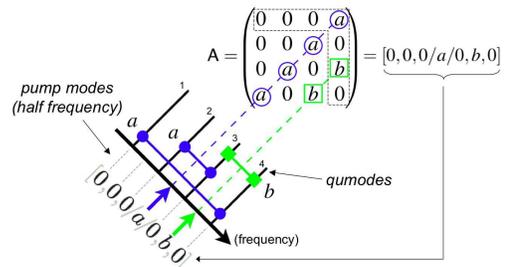}
\vspace{-.1in}
\caption{Hankel shorthand and pump specification.  A Hankel matrix is uniquely specified by the entries along the top and down the right side.  We collect these entries into a shorthand vector, representing the entire matrix itself, with the top-right entry set off with slashes.  When $\mat A$ is Hankel, its shorthand vector immediately specifies the pump spectrum required to implement $\mathcal H(\mat A)$.  Each nonzero entry in the shorthand vector denotes the amplitude of a frequency in the pump, each of which generates CV entanglement between pairs of qumodes in the OFC symmetric about half that frequency.  This accounts for all couplings prescribed by~$\mat A$.}
\label{fig:expt}
\end{center}
\vspace{-.15in}
\end{figure}

\paragraph*{QC-universal CV cluster state}

We desire a CV cluster state that is universal for one-way QC~\cite{Menicucci2006}, whose graph is bicolorable~\cite{Menicucci2007}, and whose adjacency matrix is orthogonal and Hankel for experimental simplicity~\cite{Zaidi2008}.  Orthogonality of the adjacency matrix ($\mat A \mat A^T = \mat 1$) for an undirected graph ($\mat A = \mat A^T$) yields $\mat A^2 = \mat 1$, or
\begin{align}
\label{eq:orthograph}
	(\mat A^2)_{jk} = \sum_l A_{jl} A_{lk} = \delta_{jk}\;.
\end{align}
Eq.~\eqref{eq:orthograph} has a geometric interpretation: $(\mat A^n)_{jk}$ represents the sum of the weights of all $n$-length paths from node~$j$ to node~$k$, where the weight of such an ``$n$-path'' equals the product of all edge weights along the path.  Eq.~\eqref{eq:orthograph} requires that all 2-paths that begin and end on the same node have weights that sum to~1, while those that link different nodes have weights that cancel out.

The usual candidate for a QC-universal graph is the square lattice~\cite{Raussendorf2001}.  This graph is irregular at the boundaries, so Eq.~\eqref{eq:orthograph} cannot be satisfied for any real-valued weighting of it.  Applying toroidal boundary conditions will make it regular, but orthogonality is still prohibited because there exist pairs of distinct nodes connected by only one 2-path, for which the sum in Eq.~\eqref{eq:orthograph} collapses to a single term (which must be nonzero).

\begin{figure}
\begin{center}
\includegraphics[width=.95 \columnwidth]{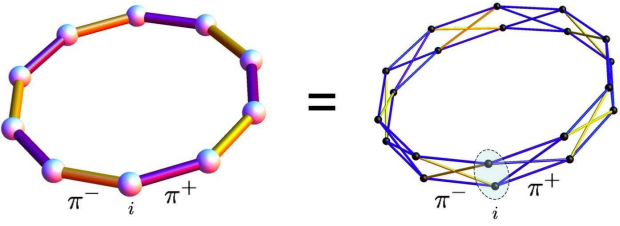}
\vspace{-.1in}
\caption{Matrix-valued weights and supergraphs. The matrix-valued weights 
$\pi^+ = \tfrac{1}{2} 
\protect\(\protect\begin{smallmatrix}
 + & + \protect\\
 + & + 
\protect\end{smallmatrix}
\protect\)$ and 
$\pi^- = \tfrac{1}{2}
\protect\(\protect\begin{smallmatrix}
 + & - \protect\\
 - & + 
\protect\end{smallmatrix}
\protect\)$, where $\pm$ stands for $\pm 1$, connect the macronodes of a ``ring" supergraph (left).  The entries in $\pi^\pm$ specify the real-valued weights in the actual ``crown" graph (right) that connects the underlying physical nodes.  Measuring $q$ for each of the physical nodes in the top layer of the crown leaves the bottom layer in a uniformly weighted ring-graph CV cluster state.}
\label{fig:matrixweight}
\end{center}
\vspace{-.15in}
\end{figure}

While real-valued weights cannot satisfy Eq.~\eqref{eq:orthograph} for a toroidal square lattice,  {\em matrix-valued weights\/} can.  In such a case, Eq.~\eqref{eq:orthograph} becomes $\sum_l \mat A_{jl} \mat A_{lk} = \delta_{jk} \mat 1$, where the ``entries''~$\mat A_{jl}$ are themselves $m \times m$ matrices.  This means $\mat A$ is now an adjacency matrix on an $(mN)$-node graph.  On the other hand, treating the $m \times m$ blocks as single entries, $\mat A$ is the matrix-weighted adjacency matrix for what we call a {\em supergraph\/}, which has {\em macronodes\/} consisting of $m$~individual nodes each.  Fig.~\ref{fig:matrixweight} illustrates the meaning of matrix-weighted edges between macronodes for the simple case of a ring supergraph, where $m=2$.

\begin{figure}[t]
\begin{center}
\includegraphics[width=.75\columnwidth]{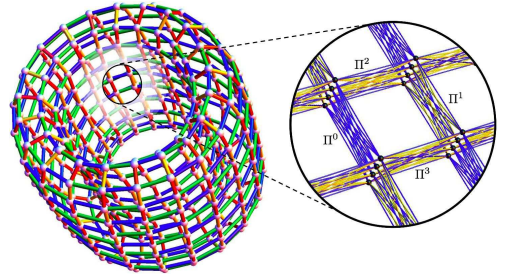}
\vspace{-.1in}
\caption{Toroidal lattice supergraph and underlying graph structure.  Each macronode in the supergraph (left) consists of four physical nodes, and each color corresponds to one of the four matrix-valued weights~$\Pi^j$ from Eq.~\protect\eqref{eq:Pidefs}.  Entries in $\Pi^j$ specify the real-valued weights connecting the underlying physical nodes (right).  Measuring $q$ on each physical node in three of the four ``layers'' leaves the remaining layer in a uniformly weighted QC-universal toroidal lattice cluster state.}
\label{fig:ttsl}
\end{center}
\vspace{-.15in}
\end{figure}

We henceforth promote the toroidal square lattice to a supergraph, illustrated in Fig.~\ref{fig:ttsl}.  The lattice has degree four, so we choose our matrix-valued weights to be four mutually orthogonal rank-one projectors onto $\reals^4$:
\begin{align}
\label{eq:Pidefs}
	\Pi^0&= \tfrac 1 4
	\left(\begin{smallmatrix}
		+ &		+ &		+ &		+ \\
		+ &		+ &		+ &		+ \\
		+ &		+ &		+ &		+ \\
		+ &		+ &		+ &		+
	\end{smallmatrix}\right)
	\;, &
	\Pi^1&= \tfrac 1 4
	\left(\begin{smallmatrix}
		+ &		- &		+ &		- \\
		- &		+ &		- &		+ \\
		+ &		- &		+ &		- \\
		- &		+ &		- &		+
	\end{smallmatrix}\right)
	\;, \nonumber \\
	\Pi^2&= \tfrac 1 4
	\left(\begin{smallmatrix}
		+ &		+ &		- &		- \\
		+ &		+ &		- &		- \\
		- &		- &		+ &		+ \\
		- &		- &		+ &		+
	\end{smallmatrix}\right)
	\;, &
	\Pi^3&= \tfrac 1 4
	\left(\begin{smallmatrix}
		+ &		- &		- &		+ \\
		- &		+ &		+ &		- \\
		- &		+ &		+ &		- \\
		+ &		- &		- &		+
	\end{smallmatrix}\right)
	\;,
\end{align}
where $\pm$ stands for $\pm 1$.  Note that the toroidal embedding in Fig.~\ref{fig:ttsl} involves a {\em twist\/} of one unit in both dimensions, a subtlety of little physical consequence but which allows for a Hankel $\mat A$ at the supergraph level (using the Hankel shorthand defined in Fig.~\ref{fig:expt}):
\begin{align}
\label{eq:4x4skewcircTTSL}
	\mat A = [\vec 0^u\!{,}\Pi^1\!{,}\vec 0^v\!{,}\Pi^0{,}\vec 0^u\!{,}\Pi^3\!{,}0{/}\Pi^2\!{/}
		\vec 0^u\!{,}\Pi^1\!{,}\vec 0^v\!{,}\Pi^0{,}\vec 0^u\!{,}{-}\Pi^3\!{,}0]\;,
\end{align}
where $\vec 0^k$ is a list of $k$ zero-matrices, the size of which is governed by context ($4 \times 4$ in this case), there are $M^2$ macronodes (with $M$ even), $u=M-1$, and $v=M^2-2M-3$.  (The negative sign on $\Pi^3$ is used, with foresight, only to make the node renumbering below work out properly.)  Orthogonality ($\mat A \mat A^T = \mat 1$) is easily verified.

The $M$-indexed family of these cluster states is universal for CV one-way QC.  To see this, first measure~$q$ on three physical nodes per macronode to reduce the supergraph to a uniformly-weighted graph with the same structure (see Fig.~\ref{fig:ttsl}).  Then cut open the toroidal lattice by measuring~$q$ along two orthogonal meridians to form an ordinary lattice, known to be universal~\cite{Menicucci2006}.

The matrix~$\mat A$, while Hankel at the supergraph level, is \emph{block-Hankel} (with $4 \times 4$ blocks) at physical node level.  A suitable renumbering of nodes will reduce the blocks to $2 \times 2$ while maintaining the block-Hankel structure:
\begin{multline}
\label{eq:2x2blockHankelTTSL}
	\mat A \cong [\vec 0^s\!{,}\pi^-\!{,}\vec 0^t\!{,}\pi^+\!{,}\vec 0^s\!{,}\pi^+\!{,}0{,}\pi^-\!{,}\vec 0^s\!{,}\pi^-\!{,}\vec 0^t\!{,}\pi^+\!{,}\vec 0^s\!{,}\pi^-\!{,}0{/}{\pi^+\!}{/} \\
	\vec 0^s\!{,}\pi^-\!{,}\vec 0^t\!{,}\pi^+\!{,}\vec 0^s\!{,}\pi^+\!{,}0{,}\pi^-\!{,}\vec 0^s\!{,}\pi^-\!{,}\vec 0^t\!{,}\pi^+\!{,}\vec 0^s\!{,}\pi^-\!{,}0]\;,
\end{multline}
where $\pi^\pm$ are defined in Fig.~\ref{fig:matrixweight}, $s = 2M-1$, $t = M^2-4M-3$, and $\cong$ indicates equality up to node renumbering.  These cluster states are \emph{QC-universal}, \emph{bicolorable}, and \emph{orthogonal}---but still only \emph{block-Hankel}.  This is nevertheless sufficient for simple experimental implementation, using a method we now describe.

\paragraph*{Experimental implementation}

In Eq.~\eqref{eq:2x2blockHankelTTSL}, each $\pi^\pm$ block corresponds to a single pump frequency.  
Such couplings can be implemented using the two orthogonal polarizations of an optical field at a given frequency, as was experimentally demonstrated by simultaneously quasiphasematching polarization-sensitive interactions ZZZ, ZYY, and YZY/YYZ (first letter is pump polarization) in a periodically poled $\mathrm{KTiOPO_4}$ (KTP) crystal~\cite{Pooser2005}.  The difference between $\pi^-$ and $\pi^+$ is thus a $180^\circ$ phase-shift in the Y-polarized pump mode.  A narrowband pump polarized at $\pm 45^\circ$ in the (ZY) plane implements a $\pi^{\pm}$ skew-diagonal band in~$\mat A$.  Equation~\eqref{eq:2x2blockHankelTTSL} therefore translates into a single OPO pumped by exactly 15~frequencies.  Interactions with OFC modes outside the desired subset must also be strictly suppressed, which can be done by cavity mirror design and/or by quasiphasematching bandwidth design~\cite{Fejer1992}. 
While the pump spectrum is relatively complex---requiring 15~frequencies---that number is {\em constant\/} with respect to the lattice size, making this construction extremely scalable.

Note that, in principle, additional physical parameters like wave vector direction or transverse mode structure could be used to directly implement block-Hankel $\mat A$'s with larger blocks (e.g., $4\times 4$).  This would reduce pump complexity but require a more sophisticated OPO.

Finally, we provide realistic estimates for the scaling potential of the CV cluster state with $N$~macronodes ($N=M^2$) and constant overall coupling strength~$\kappa$.  The pump spectrum complexity and the OFC are independent of $N$. Only two quantities scale linearly with $N$: the overall pump power and the bandwidth of the nonlinear coupling.  The number of pump photons must increase with the number of entangled qumode pairs (number of graph edges), which grows linearly with $N$ in a square lattice. Two-mode CV entanglement can be obtained at a few milliwatts pump power, with the upper limit being the optical damage limit, which in KTP is at least several watts (focused, continuous-wave), thus yielding 3 orders of magnitude of scaling range.  The bandwidth of the nonlinear coupling (100~GHz to 1~THz) must encompass the whole desired set of qumodes, separated by the cavity's free spectral range (100~MHz for a $1.5$-meter-long cavity), which yields 3 or 4 orders of magnitude.  These figures reflect ordinary---rather than state-of-the-art---performance and do not account for other interesting avenues such as implementing nonlinear couplings in slow light media. These estimates indicate our approach has a quite realistic potential for significant scaling.

\paragraph*{Finite squeezing and CV fault tolerance}
The finite-squeezing approximation is a special case of more general considerations of error correction and fault tolerance for one-way QC using CV cluster states.  Certainly, more squeezing is preferable to less, but the amount required for any particular QC task remains an open question.  Consequently, it's unclear how the nonlinear coupling strength~$\kappa$ in Eq.~\eqref{eq:Hamiltonian} will need to scale with~$N$ for any particular QC application.  Our results are nonetheless compelling because as~$N$ increases, the new interactions generated have the same squeezing strength as existing ones---i.e., existing squeezing is not ``redistributed'' to the new pairings as the lattice and pump power grow.  Moreover, scalability does enable quantum encoding redundancy. While finite squeezing errors can be mitigated in medium-sized proof-of-principle experiments with CV cluster states~\cite{Menicucci2006}, and some work also addresses CV error correction in general~\cite{Braunstein1998a, Gottesman2001}, the task of rigorously establishing a fault tolerance threshold for CV one-way QC is an important open problem.  We hope to spur further investigations along these lines. 

\paragraph*{Conclusion}
We have presented a theoretical breakthrough that opens the door to large-scale experimental generation of a universal one-way quantum computing resource, using existing technology.  The entangled states produced by this method will also be objects of interest in the study of entanglement at mesoscopic scales.  Further simplification of the method may be possible using additional degrees of freedom, such as spatial modes.

This method of entangling---in one fell swoop---a large optical frequency comb into a continuous-variable cluster state is the first ``top-down'' approach proposed for one-way quantum computation using optical encodings of information in either discrete or continuous variables.  While open questions remain about the effects of finite squeezing on scalability for particular quantum computing tasks, the unprecedented scalability of this method encourages further theoretical research. Experimental implementation is already underway.


\paragraph*{Acknowledgments}
We thank Carl Caves and Tom Gallagher for comments on the manuscript. OP was supported by National Science Foundation grants No.\ PHY-0555522 and No.\ CCF-0622100. NCM was supported by the National Science Foundation. OP and NCM were supported in part by the Perimeter Institute for Theoretical Physics, which they thank for hospitality during stimulating visits.  STF was supported by the Perimeter Institute for Theoretical Physics.  Research at Perimeter is supported by the Government of Canada through Industry Canada and by the Province of Ontario through the Ministry of Research~\& Innovation.


\vspace{-1em}
\bibliographystyle{apsrev}

\end{document}